\documentclass{article}
\usepackage{frascatiphys,here,graphicx,subfigure}
\begin{document}
\title{Self-consistent coupled-channel approach to \\
$D$ and $\bar D$ in hot dense matter 
}
\author{
Laura Tol\'os \\
{\em FIAS. J.W.Goethe-Universit\"at.} \\
{\em Ruth-Moufang-Str. 1, 60438 Frankfurt (M), Germany } \\
Angels Ramos \\
{\em Dept. d'Estructura i Constituents de la Mat\`eria. Universitat de Barcelona.}\\
{\em Diagonal 647, 08028 Barcelona, Spain }\\
Tetsuro Mizutani \\
{\em Department of Physics, Virginia Polytechnic Institute and State
University.}\\
{\em Blacksburg, VA 24061, USA }
}
\maketitle
\baselineskip=11.6pt
\begin{abstract}

A self-consistent coupled-channel approach is used to study the  properties of
$D$ and $\bar {D}$ mesons  in hot dense matter. The starting point is a broken
SU(4) $s$-wave  Tomozawa-Weinberg $DN$ (${\bar D}N$) interaction supplemented
by an attractive isoscalar-scalar term. The Pauli blocking effects, baryon
mean-field bindings, and $\pi$ and open-charm meson self-energies are
incorporated in dense matter at finite temperature. In the
$DN$ sector, the dynamically generated $\tilde\Lambda_c$ and
$\tilde\Sigma_c$ resonances
remain close to their free space position while acquiring a remarkable width
because of the thermal smearing of Pauli blocking. Therefore, the $D$ meson
spectral density shows a single pronounced quasiparticle peak close to the 
free mass, that
broadens with increasing density, and  a low energy tail associated to smeared
$\tilde\Lambda_c N^{-1}$, $\tilde\Sigma_c N^{-1}$ configurations. In the $\bar
D N$ case, the low-density approximation to the repulsive $\bar D$ self-energy
is found unreliable already at subsaturation densities.
From this study we speculate the possible  existence of $D$-mesic nuclei. We also discuss the consequences for $J/\Psi$ suppression at FAIR.

\end{abstract}
\baselineskip=14pt

Over the last years there has been a growing interest in the open ($D$, $\bar
D$) and hidden (e.g. $J/\Psi$) charmed mesons within the context of
relativistic nucleus-nucleus collisions. In particular,  $J/\Psi$ supression
was predicted as a clear signature of the formation of the quark-gluon plasma
(QGP)  \cite{MAT86}.

The future CBM (Compressed Baryon Matter) experiment of the FAIR (Facility for Antiproton and Ion Research) project at GSI will investigate, among others, the possible modifications of the properties of open and hidden charmed  mesons in a hot and dense baryonic environment. The in-medium modification of the $D (\bar D)$ mesons may explain the $J/\Psi$ suppression  in an hadronic environment, on the basis of a mass reduction of $D (\bar D)$ in the nuclear medium \cite{TSU99,MED}. Furthermore, this reduction in mass was thought to provoke  possible $D^0, D^-, \bar  {D^o}$ bound states in heavy nuclei such as $Pb$ \cite{TSU99}.

However, a self-consistent coupled-channel meson-baryon approach in nuclear medium is found essential due to the strong coupling among the $DN$ and other meson-baryon channels \cite{TOL04,TOL06,LUT06,MIZ06}, which induces the appearance of dynamically-generated resonances close to threshold.

 In the present article, we pursue a coupled-channel study of the spectral
properties of $D$ and $\bar D$ mesons in nuclear matter at finite temperatures
by extending the result of Ref.~\cite{MIZ06} to finite temperature. Then our
finding is used  to discuss the possibility of $D$-mesic nuclei as well as to examine the possible implications in the $J/\Psi$ suppression at FAIR.

\section*{Self-consistent coupled-channel approach for $D$ and $\bar D$ mesons}

 The $D$ and $\bar D$ self-energies at finite temperature are obtained from a self-consistent coupled-channel calculation taking, as bare interaction ($V$), a type of broken SU(4) $s$-wave Tomozawa-Weinberg (TW) interaction supplemented by an attractive 
 isoscalar-scalar term ($\Sigma_{DN}$). The multi-channel transition matrix $T$ 
\begin{equation}
T=V + V \ G \ T
\end{equation}
is solved using a cutoff regularization \cite{MIZ06}, which is fixed by reproducing the position and the width of 
the $I=0$ $\Lambda_c(2593)$ resonance. As a result, a new $\Sigma_c$ resonance
in the $I=1$ channel is generated around 2800 MeV.

The in-medium solution at finite temperature is obtained by
incorporating the corresponding medium modifications in the loop
function $G$. We incorporate Pauli blocking effects,  mean-field bindings of baryons  via a temperature-dependent 
$\sigma -\omega$ model, and $\pi$ and open-charm meson self-energies in the intermediate propagators (see Ref.~\cite{TOL07}). 

The $D$ ($\bar D$) self-energy is obtained self-consistently summing 
the in-medium $T_{D(\bar D)N}$ amplitudes over the thermal nucleon Fermi
distribution, $n(\vec{q},T)$, as 
\begin{eqnarray}
\Pi_{D(\bar D)}(q_0,{\vec q},T)= \int \frac{d^3p}{(2\pi)^3}\, n(\vec{p},T) \,
[{T}^{(I=0)}_{D(\bar D)N}(P_0,\vec{P},T) +
3{T}^{(I=1)}_{D(\bar D)N}(P_0,\vec{P},T)]\ , \label{eq:selfd}
\end{eqnarray}
where $P_0=q_0+E_N(\vec{p},T)$ and $\vec{P}=\vec{q}+\vec{p}$ are
the total energy and momentum of the $D(\bar D)N$ pair in the nuclear
matter rest frame, and the values ($q_0$,$\vec{q}\,$) and ($E_N$,$\vec{p}$\,) stand  for
the energy and momentum of the $D(\bar D)$ meson and nucleon, respectively, also in this
frame. The in-medium spectral density then reads
\begin{equation}
S_{D(\bar D)}(q_0,{\vec q}, T)= -\frac{1}{\pi}\frac{{\rm Im}\, \Pi_{D(\bar D)}(q_0,\vec{q},T)}{\mid
q_0^2-\vec{q}\,^2-m_{D(\bar D)}^2- \Pi_{D(\bar D)}(q_0,\vec{q},T) \mid^2 } \ .
\label{eq:spec}
\end{equation}

\section*{Open charm in hot dense matter}

\begin{figure}[htb]
\begin{center}
\includegraphics[width=0.65\textwidth]{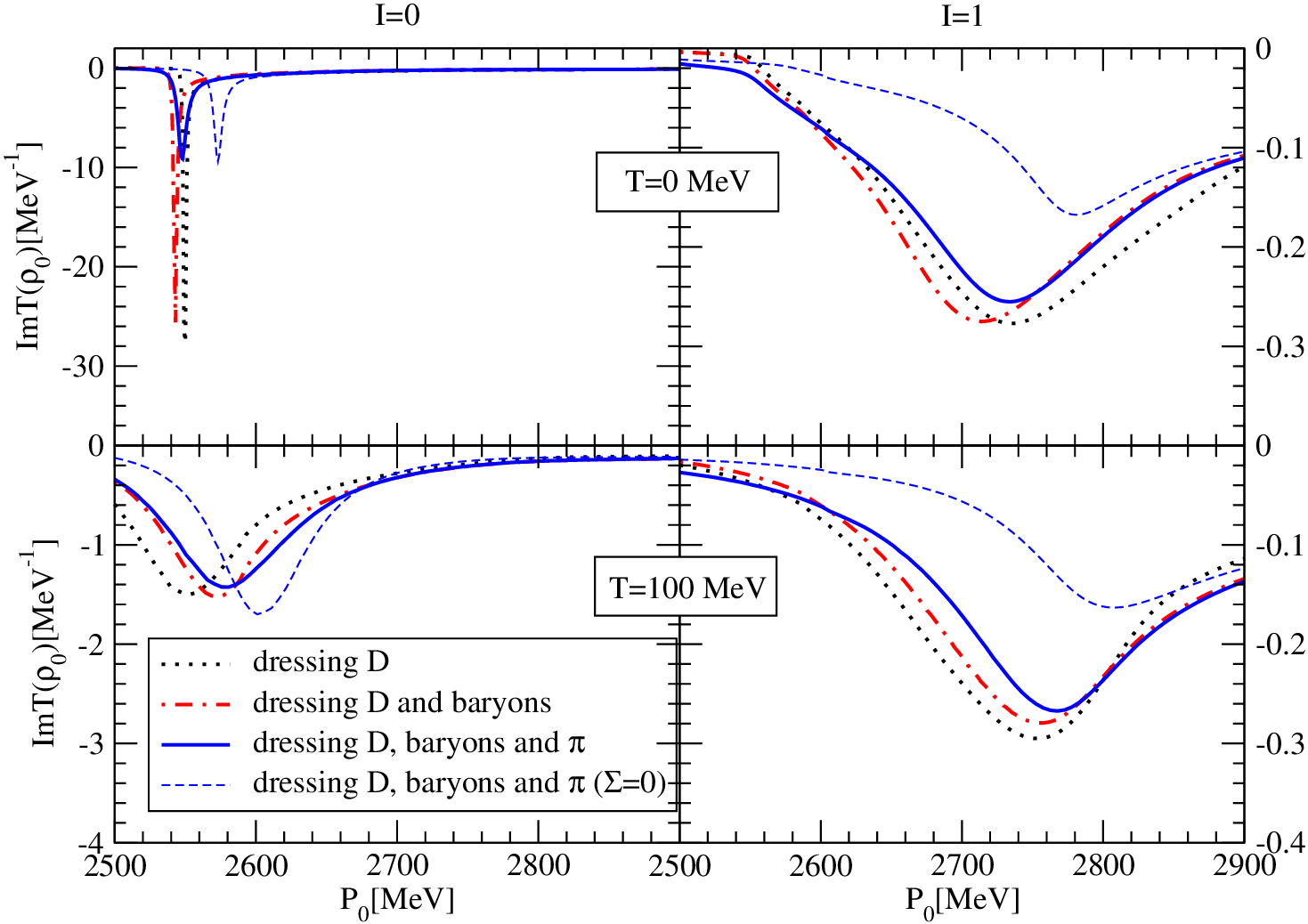}
\caption{$I=0$ $\tilde\Lambda_c$ and $I=1$ $\tilde\Sigma_c$ resonances in a hot
medium.}
\label{fig:amplituds}
\end{center}
\end{figure}

Fig.~\ref{fig:amplituds} shows the  behavior of the in-medium  $I=0$
$\Lambda_c(2593)$ and $I=1$ $\Sigma_c(2770)$ resonances, denoted as $\tilde\Lambda_c$ and $\tilde\Sigma_c$, respectively, for three different self-consistent
calculations:  i)  the self-consistent dressing of $D$ mesons only (dotted
lines), ii) adding the mean-field binding of baryons (MFB) (dash-dotted lines)
and iii) including MFB and the pion self-energy dressing (PD) (solid lines). We consider
two models: the thick lines correspond to model A ($\Sigma_{DN} \neq 0$) while
the thin-dashed lines refer to case (iii) within model B ($\Sigma_{DN}=0$).

\begin{figure}[htb]
\begin{center}
\includegraphics[width=0.6\textwidth]{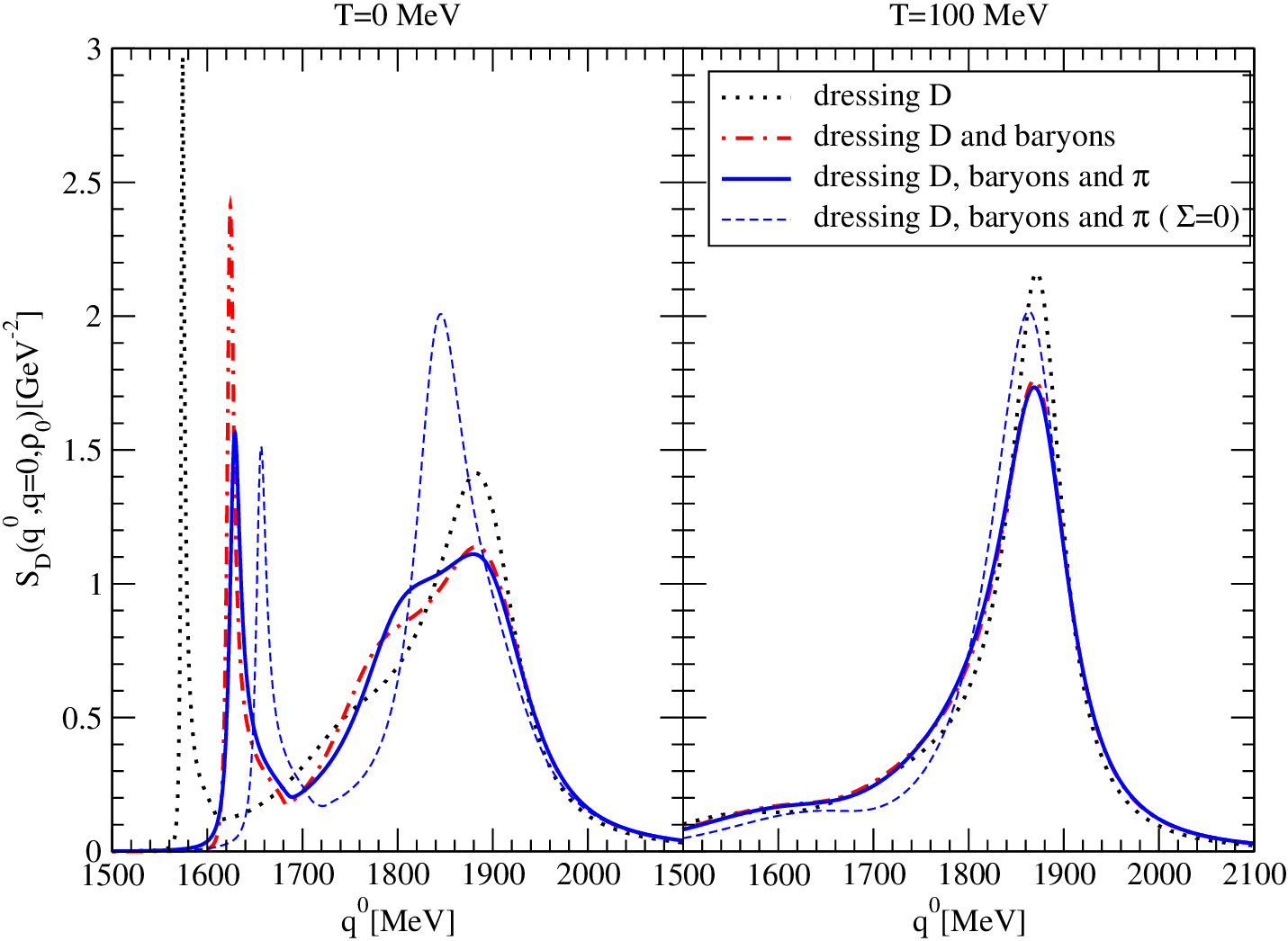}
\caption{The $q=0$ $D$ meson spectral function at $\rho_0$ for $T=0, 100$ MeV} 
\label{fig:Dspectral}
\end{center}
\end{figure}

The inclusion of medium modifications at $T=0$ lowers the
position of the $\tilde\Lambda_c$ and $\tilde\Sigma_c$ resonances with respect
to their free  values, in particular with the inclusion of MFB. Their widths, which increase due to $\tilde
Y_c N \rightarrow \pi N \Lambda_c, \pi N \Sigma_c$  processes, differ according to the phase space available. In contrast to the $\bar K N$ results \cite{RAM00,TOL02}, the PD induces a small effect in the resonances because of reduced  charm-exchange channel couplings. Still it is seen in the positions and widths through the
absorption of these resonances by one and two nucleon processes when the pion self-energy is incorporated. On the other hand, models A and B show qualitatively similar
features.

 Finite temperature results in the reduction of the Pauli blocking effects 
 due to the smearing of the Fermi surface. Both resonances move up in energy 
 closer to their free position while they are smoothen out, as seen
 previously in \cite{TOL06}.  At
$T=100$ MeV, $\tilde \Lambda_c$ is  at 2579 MeV and
$\tilde \Sigma_c$ at 2767 MeV for model A, while model B generates both
resonances  at higher energies: $\tilde \Lambda_c$ at 2602 MeV and
$\tilde \Sigma_c$ at 2807 MeV.

\begin{figure}[htb]
\begin{center}
\includegraphics[width=0.65\textwidth]{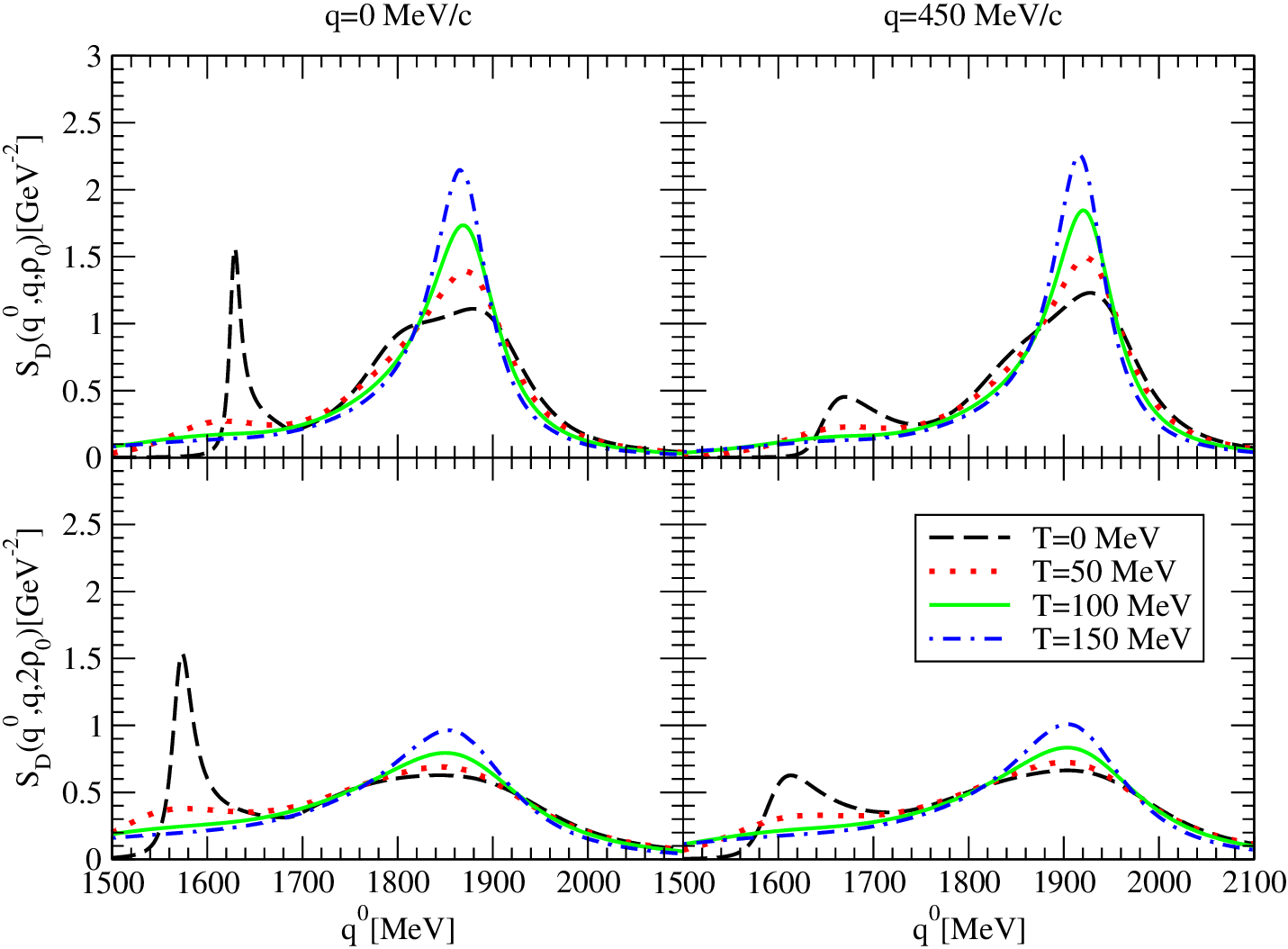}
\caption{The evolution of the $D$ meson spectral function with temperature} 
\label{fig:spec_evol}
\end{center}
\end{figure}

\begin{figure}[!ht]
\begin{center}
\includegraphics[width=0.45\textwidth]{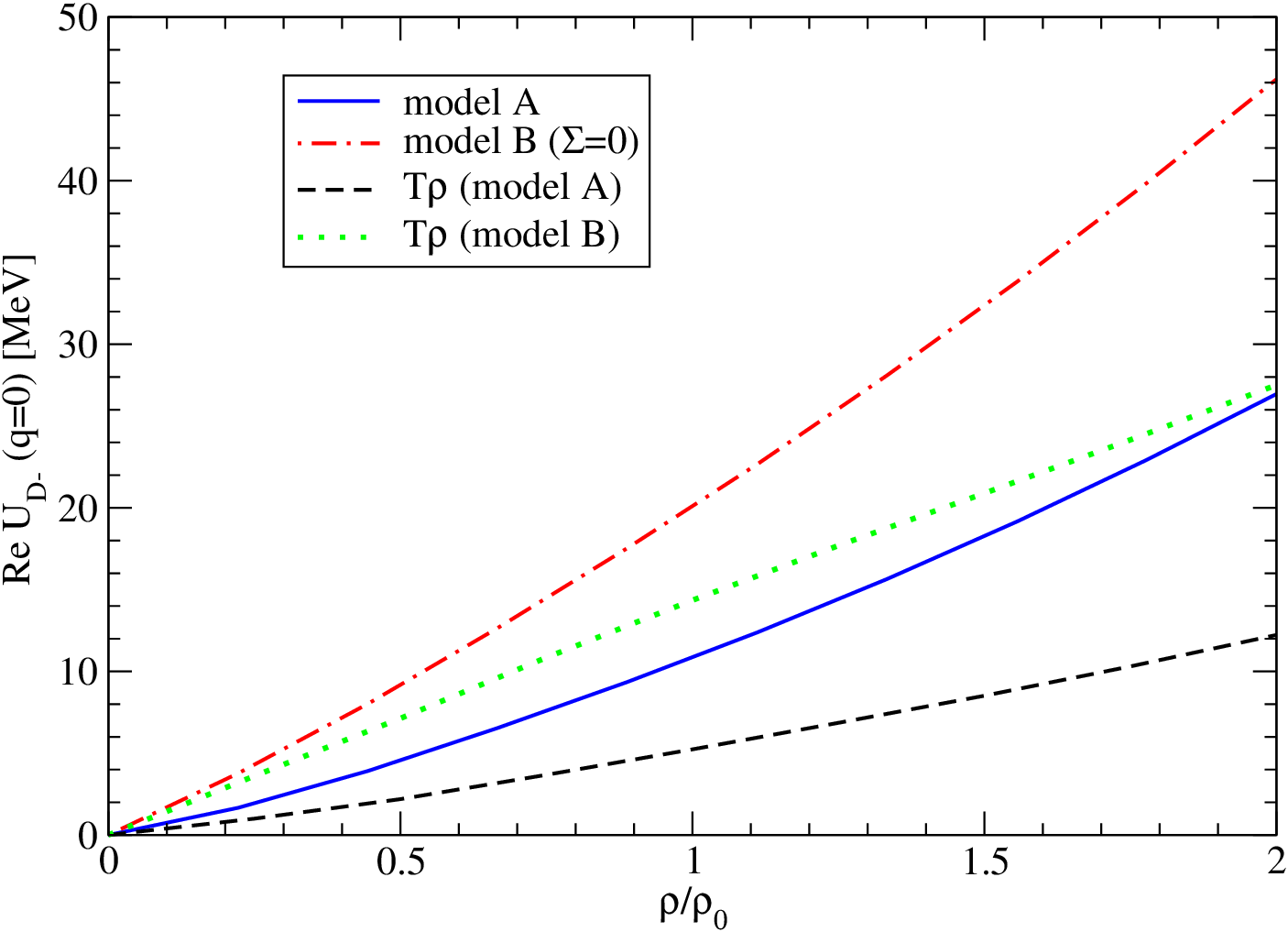}
\caption{$\bar D$ mass shift and the low-density approximation }
\label{fig:Dmass}
\end{center}
\end{figure}

\begin{figure}[htb]
\begin{center}
\includegraphics[width=0.65\textwidth]{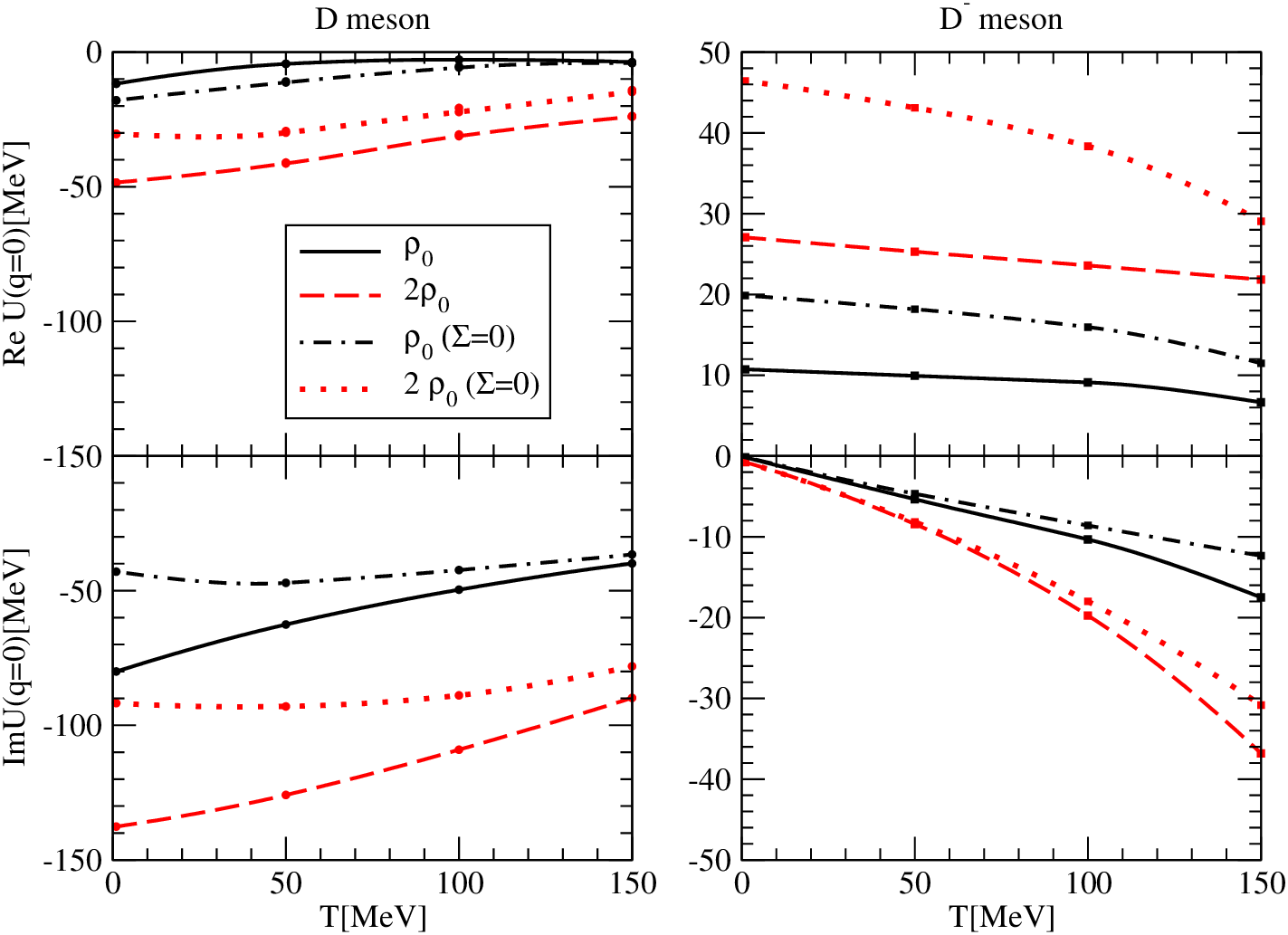}
\caption{The $D$ and $\bar D$ potentials as function of temperature} \label{fig:DDbar}
\end{center}
\end{figure}

We display in Fig.~\ref{fig:Dspectral} the $D$ meson spectral function at zero
momentum obtained in cases (i) to (iii) for model A (thick lines) and in 
case (iii) for model B
(thin line) at saturation density, $\rho_0=0.17 \ {\rm fm^{-3}}$. Two peaks
appear in the spectral density at $T=0$. The lower one corresponds to the
$\tilde \Lambda_c N^{-1}$ excitations,
whereas the  higher one is the
quasi(D)-particle  peak  mixed with  the $\tilde \Sigma_c N^{-1}$ state. The
lower energy mode goes up by about $50$
MeV relative to (i) when MFB effects are included. The reason is that the meson
requires to carry more energy to excite the $\tilde \Lambda_c$ in order to
compensate for the attraction felt by the
nucleon. The same effect is observed for the $\tilde \Sigma_c N^{-1}$
configuration that mixes with the quasiparticle peak. As expected, the PD does
not alter much the position of $\tilde \Lambda_c N^{-1}$ excitation or the
quasiparticle peak. For model B (case (iii) only), the absence of the
$\Sigma_{DN}$ term moves the $\tilde \Lambda_c N^{-1}$ excitation closer to
the  quasiparticle peak, while the latter completely mixes with the $\tilde
\Sigma_c N^{-1}$ excitation. 

Finite temperature effects result in the dilution of those structures with
increasing temperature while the quasiparticle peak gets closer to its free
value and it becomes narrower. This is due to the fact that the self-energy
receives contributions from $DN$ pairs at higher momentum where the interaction is weaker.

The evolution of the spectral function with temperature is seen in
Fig.~\ref{fig:spec_evol} for two densities, $\rho_0$ and $2\rho_0$,
and two momenta, $q=0$ MeV/c and $q=450$ MeV/c, in case (iii) for model A. As
in the previous figure, we observe how the $\tilde \Lambda_c
N^{-1}$ and $\tilde \Sigma_c N^{-1}$  structures dissolve
with increasing temperature,
while the quasiparticle peak becomes narrower and moves closer to its free
value position. The widening of the quasiparticle peak for larger nuclear
density may be understood as due to enhancement of collision and absorption
processes. The
$\tilde\Lambda_c N^{-1}$ mode moves down in energy
with increasing density due to the lowering in the position of the $\tilde\Lambda_c$
resonance induced by the more attractive $\Sigma_{DN}$ term in model A.

With regard to the $\bar D N$ sector, we first study the effective $\bar D N$
interaction in free space and, in particular, the $\bar D N$ scattering lengths.
 For model A (B) those are $a^{I=0}=0.61 \ (0)$ fm and   $a^{I=1}=-0.26
\ (-0.29)$ fm.   The zero value of the $I=0$ scattering length for model B is
due to the vanishing coupling coefficient of the corresponding pure TW  $\bar
DN$ interaction. This is in contrast to the repulsive $I=0$ scattering length
reported in \cite{LUT06}, while agreement is found in the $I=1$ contribution.
In the case of model A, the non-zero value of the $I=0$ scattering length is
due to the magnitude of the $\Sigma_{DN}$ term. Our results are consistent with
those of a recent calculation based on meson- and one-gluon exchanges
\cite{HAI07}.

Next, we display the $\bar D$ mass shift in cold nuclear matter in
Fig.~\ref{fig:Dmass}.  For both A and B models, the mass shift is repulsive due to the
$I=1$ dominant component. However, despite the
absence of resonances in the $\bar D N$ interaction, the low-density 
$T \rho$ approximation 
breaks down at relatively low densities, so it is not applicable at saturation
density.

The comparison between $D$ and $\bar D$ optical potentials at $q=0$ MeV/c as
functions of temperature for two different densities ($\rho_0$ and $2\rho_0$)
is shown in Fig.~\ref{fig:DDbar}. For model A (B) at $T=0$ and $\rho_0$, the
$D$ meson  feels an attractive potential of $-12$ ($-18$) MeV  while the $\bar
D$ feels a repulsion of $11$ ($20$) MeV. A similar shift for $D$
meson mass is obtained in Ref.~\cite{TOL06}. The temperature dependence of the
repulsive real part of the ${\bar D}$ optical potential is very weak, while the
imaginary part increases steadily due to the increase in the collisional width. The
picture is somewhat different for the $D$ meson due to the overlap of the
quasiparticle peak with the $\tilde{\Sigma}_c N^{-1}$ mode. Furthermore, the
in-medium behavior of the $\tilde{\Sigma}_c N^{-1}$ mode is determinant for
understanding the effect of the $\Sigma_{DN}$ term on the $D$ meson potential.

Taking into account our results, we might look at the question of possible
$\bar{D}$ bound states discussed in \cite{TSU99}. While $D^-$ -mesic atoms
can always be bound by the Coulomb interaction, no strongly bound
nuclear states or even bound $\bar{D}^0$ nuclear systems are expected due to
the repulsive ${\bar D}$-nucleus optical potential at zero momentum. 
In the charm $C=1$ sector,
an experimental observation of bound $D$ nuclear states is ruled
out by the moderate attraction and large width found for the $D$ meson optical
potential.
 
With respect to the $J/\Psi$ suppression, the in-medium $\bar D$ mass is seen
to  increase by about $10-20$ MeV whereas the tail of the quasiparticle peak of
the $D$ spectral function extends to lower "mass" due to the thermally spread
$\tilde Y_c N^{-1}$. Nevertheless, it is very unlikely that this lower tail
extends as far down by 600 MeV with sufficient strength to influence the
$J/\Psi$ production .  So the only way for the $J/\Psi$ suppression to take
place is by cutting its supply from the excited charmonia: $\chi_{c\ell}(1P)$
or $\Psi'$, which will be strongly absorbed in the medium
by multi-nucleon processes.

\section{Acknowledgments}

This work is partly supported by contracts
MRTN-CT-2006-035482 (FLAVIAnet, EU), RII3-CT-2004-506078 (HadronPhysics, EU),
by FIS2005-03142 (MEC, Spain), 2005SGR-00343 (Generalitat de Catalunya),
ANBest-P and BNBest-BMBF 98/NKBF98 (Germany).

\end{document}